\documentclass[12pt,singlecolumn]{emulateapj}

\usepackage{latexsym}
\usepackage{mathrsfs}
\usepackage{times} 
\usepackage{graphics}
\usepackage[german,english,american]{babel} 
\newcommand{\sci}{Science}
\newcommand{\cirx}{Circinus~\mbox{X-1}}

\begin{document}

\shorttitle{The X-ray jet of Circinus X-1}
\shortauthors{Heinz, Schulz, Brandt, \& Galloway}
\title{Evidence for a parsec scale X-ray jet from the accreting
neutron star Circinus X-1}

\author{S.~Heinz\altaffilmark{1}, N.~S.~Schulz\altaffilmark{2},
 W.~N.~Brandt\altaffilmark{3}, \& D.~K.~Galloway\altaffilmark{4,5}}
 \altaffiltext{1}{Astronomy Department, University of
 Wisconsin-Madison, 475 N. Charter St., Madison, WI 53706; {\em
 heinzs@astro.wisc.edu}}\altaffiltext{2}{Massachusetts Institute of
 Technology, 77 Massechusetts Ave., Cambridge, MA
 02139}\altaffiltext{3}{Department of Astronomy and Astrophysics, Penn
 State University, 525 Davey Lab, University Park, PA
 16802}\altaffiltext{4}{School of Physics,
 University of Melbourne, Parkvile, Victoria 3010,
 Australia}\altaffiltext{5}{Centenary Fellow}

\begin{abstract}
We analyzed the zero-order image of a 50 ks {\em Chandra} gratings
observation of \cirx, taken in 2005 during the source's low-flux
state. \cirx~is an accreting neutron star that exhibits
ultra-relativistic arcsecond-scale radio jets and diffuse
arcminute-scale radio jets and lobes.  The image shows a clear excess
along the general direction of the north-western counter-jet,
coincident with the radio emission, suggesting that it originates
either in the jet itself or in the shock the jet is driving into its
environment.  This makes {\cirx}~the first neutron star for which an
extended \mbox{X-ray} jet has been detected.  The kinetic jet power we
infer is significantly larger than the minimum power required for the
jet to inflate the large scale radio nebula.

\end{abstract}

\section{Introduction}
\label{sec:introduction}
\cirx~is an unusual, highly variable \mbox{X-ray} binary.  Beyond
variability at the 16.5 day orbital period, which includes deep dips
near orbital phase 0, its long-term light curve can reach peak fluxes
between 1 -- 2 Crab, but also shows several long incursions into a
very deep low-flux state.  The compact object in this system is
believed to be a neutron star, based on observed type I \mbox{X-ray} bursts
\citep{tennant:86}, putting the source at a likely distance of
$\approx 7.8-11 {\rm kpc}$ \citep{jonker:04,jonker:07}, roughly
consistent with the estimated hydrogen absorption column of $N_{\rm
H}\approx 2\times 10^{22}\,{\rm cm^{-2}}$ \citep[e.g.][]{schulz:02}.
We will adopt a distance of $D=7.8\,{\rm kpc}\,D_{7.8}$ throughout
this letter.

One of the remarkable characteristics of the source is its radio
structure: on arcmin scales, \cirx~shows two radio jets
(running south-east to north-west), embedded in a large scale, diffuse
radio nebula \citep[e.g.][]{stewart:93,tudose:06}.  This nebula is
most likely the radio lobe inflated by the jets over several hundred
thousand years \citep{heinz:02}.  The arcmin-scale jets are clearly
curved, a possible sign of interaction with the interstellar medium or
jet precession.  On arcsec scales, radio monitoring revealed a
one-sided, highly variable jet \citep{fender:04}, with a lower limit
on the jet Lorentz factor of 16, making \cirx~the only neutron-star
\mbox{X-ray} binary with a large-scale radio jet and, at the same time, the
fastest known microquasar.

Because relativistic jets are launched in the inner regions of
accretion flows, they are important probes of strong gravity and the
physics of compact objects.  One fundamental question in the study of
microquasars is if and how neutron star jets are different from
black-hole jets.  As the only neutron star with a known
ultra-relativistic, powerful jet, \cirx~is a very important object: in
the radio, the jets exhibit all the hallmarks one would otherwise
expect from a black hole.  Does this equivalence carry over to other
wave bands? The discovery of resolved \mbox{X-ray} jets from a number of
black-hole \mbox{X-ray} binaries \citealt{corbel:02,corbel:05} suggests that
{\em Chandra} might be able to detect extended jet emission from \cirx~as well.

\cirx~is overwhelmingly \mbox{X-ray} bright during its high state, rendering
any extended jet emission unobservable. Only in its \mbox{low-flux}
state could we hope to detect this emission.  In \S\ref{sec:data},
we analyze the zero-order image of a {\em Chandra} observation during
this state and argue that the extended excess emission we detect
originates in the counter-jet, \S\ref{sec:discussion} discusses the
implications.

\section{Observations and Analysis}
\label{sec:data}

\cirx~was observed with the high energy transmission gratings (HETGS;
\citealt{canizares:05}) onboard the {\em Chandra} \mbox{X-ray} Observatory on
June 2nd 2005 (04:13:01 UT, OBSID 5478) as part of a campaign to
obtain high-resolution \mbox{X-ray} spectra of the source during its
extremely low long-term \mbox{X-ray} flux state. The observation lasted 50 ks
and occurred at orbital phases 0.06 -- 0.10.  As a byproduct of this
observation we obtained a zero-order image of the source.  The source
flux was exceptionally low and the point spread function (PSF) is
almost completely pileup free due to the application of a 1/2 subarray
mode which held the CCD frametime to 1.7 s.

The upper panels of Fig.~1 show the zero-order image. The
colors represent increasing count levels from black (low) to white
(high). With the HETG in place the zero-order image is affected by
various artifacts. Visible are the four dispersion arms of the
positive and negative orders of the medium and high energy
gratings. These lie at position angles of $\sim 26^{\circ},
36^{\circ}$, and the same at 180$^{\circ}$ rotation (position angles
throughout this letter are measured counter-clockwise from due West
from the point source and abbreviated as PA), outward of about 1.3
arcmin.  Furthermore the spectrometer shields some of the soft
\mbox{X-ray} photons below 0.7 keV and reduces the overall throughput in an
azimuthally uniform fashion. The zero-order PSF is otherwise not
affected by the spectrometer.  For brighter sources the CCD read out
streak (a charge trace along the PSF centroid column) becomes
prominent.  In order to enhance contrast in the PSF wing areas the
core of the PSF in Fig.~1 is over-exposed.

\subsection{The X-ray jet}
\label{sec:length}
Figure 1 shows a distinctive surface brightness excess in the
north-western quadrant that does not line up with any of the known PSF
artifacts (see white arrow).  The morphology 

{\noindent \begin{minipage}{\textwidth}
\begin{center}{\resizebox{0.96\columnwidth}{!}{\includegraphics{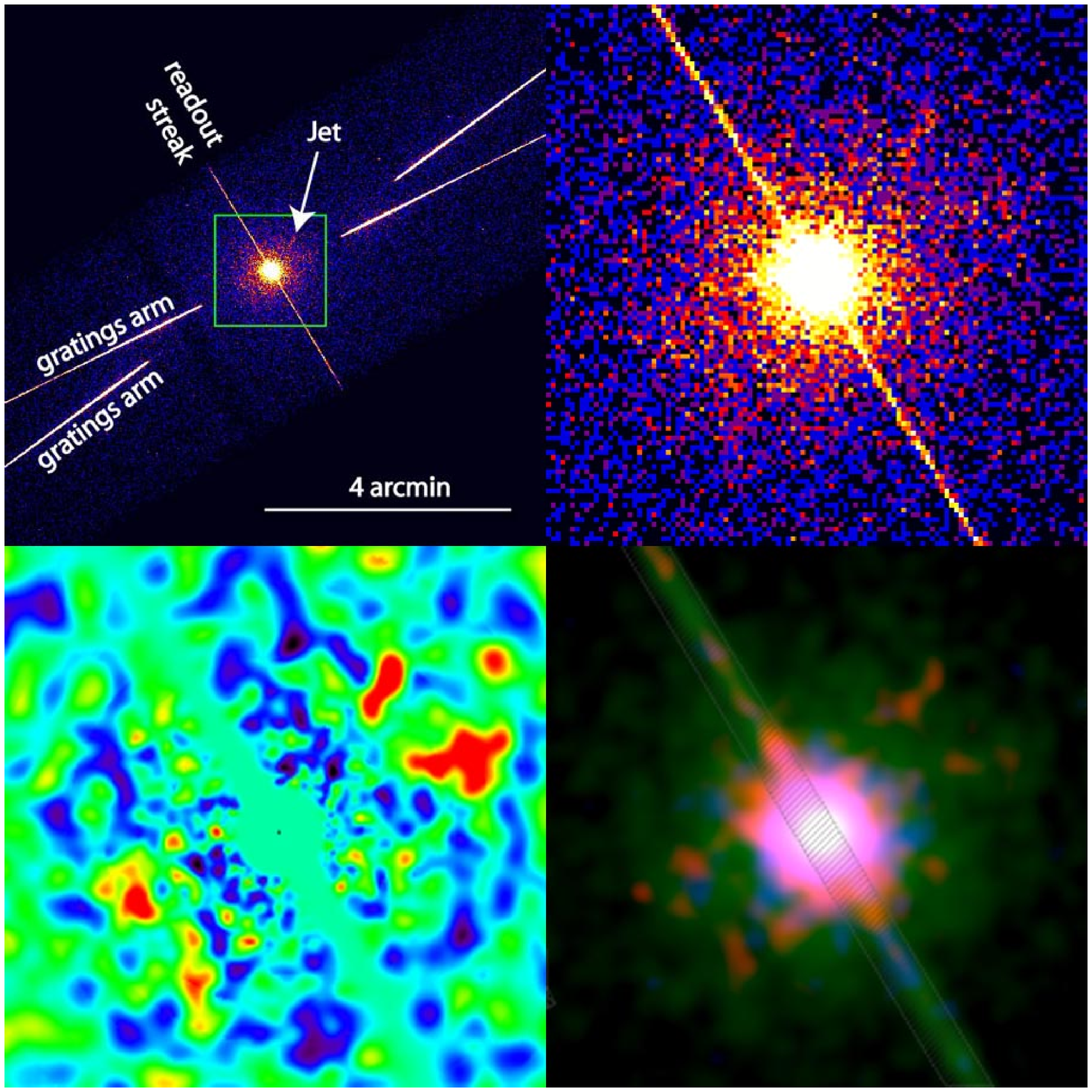}}}
\end{center}
{\small \ \ \ Fig.~1: (a) {\em top left:} 0.3-10 keV counts image of \cirx.
The white arrow denotes the position of the \mbox{X-ray} jet, the green box
is the region displayed in the other three panels; (b) {\em top
right:} enlarged image of the analysis box; (c) {\em bottom left:}
adaptively smoothed, cleaned difference image with 5$\sigma$
significance per smoothing length. Jet emission is visible in red; (d)
{\em bottom right:} smoothed color image (red: 2-4 keV, green: 4-7
keV, blue: 7-10 keV), showing that the jet is soft (red). Hatched area
shows regions affected by the read out
streak.}
\end{minipage}}
\vspace*{12pt}

{\noindent of the feature resembles two filaments pointing away from
the point source at PAs $22.5^{\circ}$ and $54.5^{\circ}$.  The bottom
panel of Fig.~2 shows a combined radial
surface-brightness profile along two 15$^{\circ}$ wide sectors
centered on those two PAs (see insert for extraction region), clearly
showing the excess emission in the region from about 0.45 to 0.9
arcmin. The formal statistical significance of the excess over the
azimuthally smoothed background emission is 7.1 sigma.  A Gaussian fit
yields a centroid distance of $r_{\rm peak} \approx 0.64\pm 0.03$
arcmin from the point source and $\sigma \approx 0.13\pm 0.02\,{\rm
arcmin}$.}

In order to illustrate the excess more clearly, we adaptively
\newline
\vspace*{7.63in} 

{\noindent smoothed the central region of the zero-order image after
removing all the gratings and CCD artifacts, subtracted the
azimuthally averaged profile $\Sigma_{\rm rad}$, and divided the
difference image by $\Sigma_{\rm rad}$ to create a normalized
difference image (panel c, Fig.~1).  The adaptive
smoothing length was chosen to vary only with radial distance from the
point source to provide an average significance of 5$\sigma$ per
smoothing length.}

Figure 2 also shows an azimuthal surface brightness profile across an
annular sector covering the excess emission. The filaments show clear
lateral extent: For comparison, Fig.~2 also shows an azimuthal profile
across the readout streak, which is representative of the image
resolution, and is much narrower.  Fitting two Gaussians to the excess
yields centroid angles of ${22.5^{\circ}}\pm 0.6^{\circ}$ and
${54.5^{\circ}}\pm 0.5^{\circ}$ and widths of $\sigma \sim
{2.3^{\circ}}\pm 0.6^{\circ}$ and $\sigma \sim {2.0^{\circ}}\pm
0.5^{\circ}$, respectively, spanning an angle of $32^{\circ}$ from
peak to peak.  Since the inclination angle $i$ of the feature (angle
to the line of sight) is unknown, this is an upper limit on the
opening angle.
\setcounter{figure}{1}

\begin{figure}[tbp]
\begin{center}{\resizebox{\columnwidth}{!}{\includegraphics{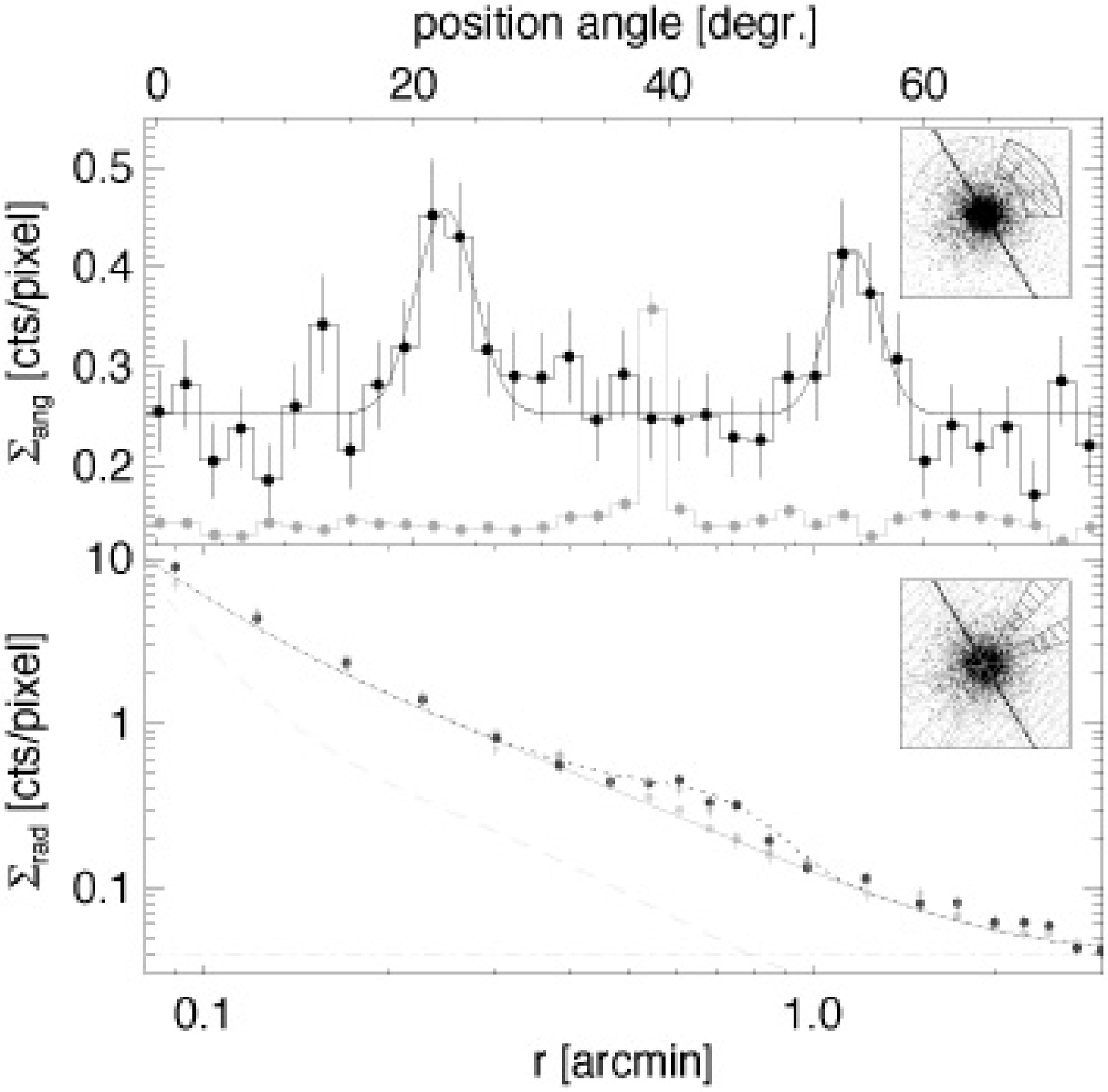}}}\end{center}
\caption{{\em Top panel:} azimuthal surface brightness profile across
the jet (black) and across the readout streak (grey, scaled and
shifted to line up with jet excess). {\em Bottom panel:} radial
surface-brightness profile across the jet (black) and azimuthal
average from regions excluding the jet, readout streak, and gratings
arms (grey). Also shown: best fit analytic approximation $\Sigma_{\rm
rad}$ (grey line) and best Gaussian fit to jet excess (black dashed
line).  Grey diamonds show the radial surface brightness profile from
the region of the approaching radio jet (offset 180$^{\circ}$ from
\mbox{X-ray} jet, centered around PA 244$^{\circ}$).  Grey dashed
line: Marx-simulated PSF. Grey dash-dotted line: off-source
background. {\em Inserts:} extraction regions relative to analysis box
from Fig.~1; black-hatched (bottom-right to top-left:
counter-jet; grey-hatched (bottom-right to top-left): jet; gery
hatched (bottom-left to top-right): background (bottom) or PSF from
streak (top). \label{fig:fig2}}
\end{figure}
\hbox{ }

It is noteworthy that even the azimuthally symmetric emission is well
in excess of the PSF estimate derived from Marx ray-tracing.  However,
calibration observations show that the ray-traced PSF underestimates
the true PSF by a factor of order 2-3, especially at high energies,
making this excess emission marginal at best.  It is possible that
such excess emission stems from the large scale synchrotron nebula
around the source.  However, given the questionable significance,
detailed discussion of this spherical excess is beyond the scope of
this letter.

\subsection{Comparison with the radio jet}
\label{sec:radio}
The excess emission is generally aligned with the north-western
arcmin-scale radio jet.  Multiple epochs of radio observations with
varying resolution exist of the \cirx~jet. \citet{stewart:93}
presented 1991 ATCA observations (12 arcsec resolution) which show
diffuse extended emission on arcmin scales and a jet-like enhancement
in the south--east to north--west direction.  On the scale of the
observed \mbox{X-ray} feature, we estimate the jet PA to be roughly
${33^{\circ}}\pm 9^{\circ}$, with significant bending towards larger
PAs on larger scales.  \citet{tudose:06} presented recent ATCA
observations on the same scales as \citet{stewart:93}.  In the highest
resolution image (2004 epoch), we estimate a PA of ${40^{\circ}}\pm
15^{\circ}$.  Finally, \citet{fender:04} presented a series of
well-resolved jet images (10/2000-12/2002) on scales of about 10
arcsec, with a PA of $\approx {-130^{\circ}}\pm 4^{\circ}$.  While
relativistic Doppler boosting hides the receding north-western jet
from view, it is reasonable to expect that it travels in the opposite
direction from the approaching jet, implying a PA of $\approx
{50^{\circ}}\pm 4^{\circ}$.  All of these angles fall between the two
\mbox{X-ray} filaments.

Figure 3 shows an overlay of the radio contours from \citet[][blue
contours]{tudose:06}, the \mbox{X-ray} image (grey scale) and the smoothed
jet contours from Fig.~1c (red contours).  Also shown are the limits
on the PA for the approaching arcsec-scale jet \citep[][green
lines]{fender:04}.  It is clear from the figure that the two
north-western \mbox{X-ray} arms are closely ``hugging'' the radio jet on
either side.  We thus interpret the features as the \mbox{X-ray} counterpart
to the receding north-western radio jet from \cirx.

The light travel time from the point source to the location of the
\mbox{X-ray} jet is $\tau_{\rm c} \approx 5 {\rm yrs}/\sin{i}$ (where the
inclination $i$ is the angle between the jet and the line of sight).
Since $i$ is not known, we cannot directly associate the \mbox{X-ray} jet
with a specific radio episode or \mbox{X-ray} state change of the binary.
Since we observed \cirx~in 2005, any associated central flare must
have happened before 2000.  This is also the time scale on which one
would expect changes to the \mbox{X-ray} jet to occur (e.g., proper motion),
which should be detectable by future {\em Chandra} observations.

\subsection{The X-ray spectrum}
\label{sec:spec}
The spectrum of the point source \cirx~has been analyzed by
\citet{schulz:07}.  The spectrum appears extremely hard in this very
low-flux state, with virtually no counts below 1 keV.  This hard
spectrum is reflected in the diffuse emission from the wings of the
PSF.  Figure 1d shows a smoothed color image of the inner
region, indicating that the jet emission has a significantly softer
\mbox{X-ray} color than the background (red: 1-4 keV, green: 4-7 keV, blue:
7-10 keV).  Note that the other region with visibly softer spectrum is
the south-eastern quadrant, close to where we would expect to find the
approaching jet (see \S\ref{sec:apjet} for a brief discussion of the
south-eastern emission).

\begin{figure}[tbp]
\begin{center}{\resizebox{\columnwidth}{!}{\includegraphics{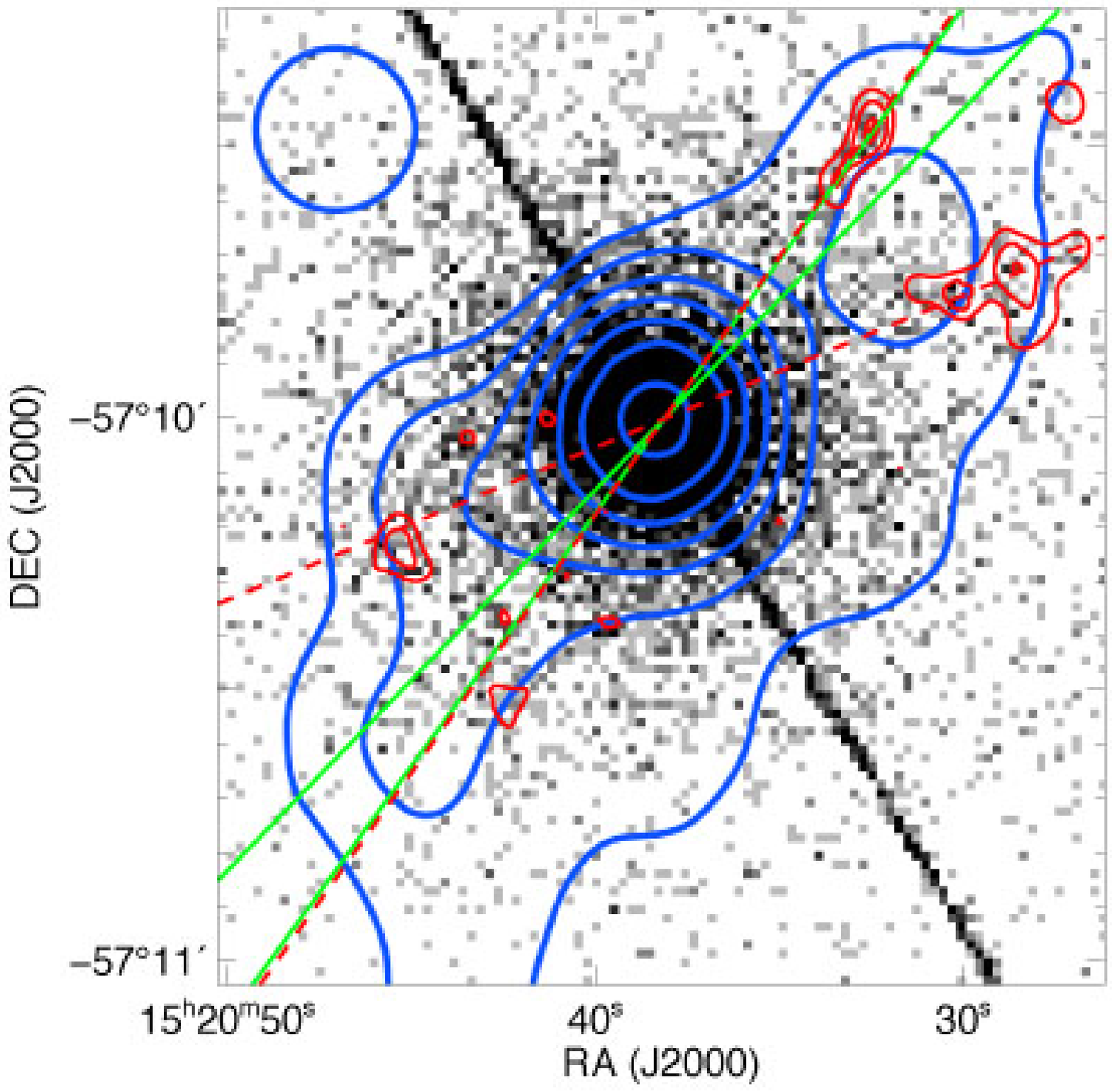}}}\end{center}
\caption{Radio-X-ray overlay. Blue contours: 1.4\,GHz surface
  brightness \citep[adapted from][levels increase by $\sqrt{2}$
  between contours; outermost contour: 22 mJy/beam; beam size shown on
  top left]{tudose:06}; grey scales: \mbox{X-ray} image (Fig.~1b); red
  contours: adaptively smoothed, normalized, PSF-subtracted image
  (Fig.~1c); green lines: estimated allowed range of PAs from
  high-resolution radio observations of approaching jet
  \citep{fender:04}; red lines: allowed range in PA for \mbox{X-ray} jet from
  Fig.~2 (top panel).\label{fig:fig3}}
\end{figure}

The background-subtracted count rate from the jet is $3.5 \times
10^{-3}\,{\rm s^{-1}}$, giving roughly 175 counts from the jet over
480 background counts.  Given the limited number of counts, we
restricted fitting to the simplest possible models. The first model we
fitted was an absorbed power-law, giving $N_{\rm H} \approx
5.9^{+6.9}_{-3.0}\times 10^{22}\,{\rm cm^{-2}}$ and $\Gamma \approx
{3.0}^{+2.6}_{-1.5}$.  As it stands, the best-fit photon index is
broadly consistent with the emission being of synchrotron origin (for
which we would expect $1.5 \lesssim \Gamma < 2.5$).  An absorbed
thermal model can fit the data equally well, giving $N_{\rm H}
5.4^{+5.7}_{-3.2} \times 10^{22}\,{\rm cm^{-2}}$ and
$T=2.2^{+7.0}_{-1.1}$ keV.

Taking the best-fit spectral parameters at face value, we can estimate
the un-absorbed 2-10 keV source flux to be roughly $F_{\rm
2-10}\approx 1.7\,\times 10^{-13} {\rm ergs\,cm^{-2}\,s^{-1}}$,
corresponding to an \mbox{X-ray} luminosity of $L_{\rm 2-10} \approx
1.2\times 10^{33}\,{\rm ergs\,s^{-1}}\,D_{7.8}^2$.

\section{Discussion}
\label{sec:discussion}
\subsection{Interpretation and physical parameters}
\label{sec:synchrotron}
The V-shaped morphology of the \mbox{X-ray} jet suggests
limb-brightened emission from the surface of a conical volume.  The
length of the \mbox{X-ray} jet is roughly $l_{\rm jet} \approx
1.6\,{\rm pc}\,D_{7.8}/\sin{i}$.  Given the half-opening angle of
roughly $\alpha \sim 2.6^{\circ}\sin{i}$, the volume of the emitting
cone is $V_{\rm jet} \approx 1.2\times 10^{55}\,{\rm
cm^3}\,D_{7.8}^3/\sin{i}$.  Since we can observe the radio and
\mbox{X-ray} counter-jet on arcmin scales, it is safe to assume that
its motion is no longer ultra-relativistic and we can neglect Doppler
corrections.  We propose two possible alternative explanations for the
origin of the excess emission:

{\noindent {\bf Synchrotron emission from the jet:} The \mbox{X-ray}
emission could be limb-brightened emission from the conical walls of
the jet itself.  Assuming a synchrotron slope of $\Gamma \approx 1.5$
and a volume filling fraction of order unity, the equipartition
particle pressure is $p_{\rm eq} \approx 5\times 10^{-12}\,{\rm
ergs\,cm^{-3}}\,D_{7.8}^{-4/7}\,(\sin{i})^{4/7}$.  We can then
estimate the minimum jet power in the counter-jet to be roughly}
$W_{\rm jet} \gtrsim 5\times 10^{36}\,{\rm
ergs\,s^{-1}}\,D_{7.8}^{10/7}\,(\sin{i})^{4/7}$ which is larger than,
and thus consistent with, the lower limit on the time-averaged kinetic
power $\langle W \rangle \gtrsim 10^{35}\,{\rm ergs\,s^{-1}}$ from the
large scale radio nebula \citep{heinz:02,tudose:06}.
\vspace*{3pt}
	
{\noindent {\bf Thermal emission:} The \mbox{X-rays} could also be due
to thermal emission from the shock driven into the ISM by the
propagation of the jet alongside the expanding cocoon.  In this case,
the required electron density would be $n_{\rm thermal} \approx
10\,{\rm cm^{-3}}\,D_{7.8}^{-1/2}\,(\sin{i})^{1/2}$, with a total
emitting gas mass of $M_{\rm thermal} \approx
0.1\,M_{\odot}\,D_{7.8}^{5/2}/(\sin{i})$ and a very high pressure of
$p_{\rm thermal} \approx 3\times 10^{-8}$.  Since the shock velocity
is fixed by the temperature, we can estimate the jet power required to
explain the X-rays to be $W_{\rm thermal} \approx 5\times
10^{36}\,{\rm ergs\,s^{-1}}\,D_{7.8}^{3/2}\,(\sin{i})^{1/2}$, which is
similar to the power estimated for the synchrotron case, making
$W_{\rm kin}$ a relatively robust estimate.
\vspace*{3pt}

{\noindent {\bf Inverse Compton emission:} In order for inverse
Compton emission to explain the \mbox{X-ray} jet, potential seed photons
would have to come from the diffuse radio lobe, from the radio jet (as
synchrotron-self Compton), or from the binary.  In all three cases,
the pressure required to produce the observed \mbox{X-ray} flux falls into
the range $p_{\rm IC} \sim 0.001$ to $1 \,{\rm ergs\,cm^{-3}}$, which
can be ruled out on energetic grounds. Thus, inverse Compton
scattering is not a viable radiation mechanism.}

\subsection{The south-eastern X-ray jet}
\label{sec:apjet}
The radial surface brightness profile along the direction of the
approaching jet in Fig.~2 shows a small excess over the average
profile is visible from the plot, significant at the 4$\sigma$ level.
Given that the approaching jet is observed in the radio, the low
intensity of the approaching \mbox{X-ray} jet is noteworthy and could
have one or both of the following causes:

{\noindent {\bf a)} The dimness of the approaching jet could reflect
  differences in the environment on the near and far side of the
  source.  Given the clear non-spherical appearance of the radio
  nebula, such an asymmetry in ISM density and/or pressure is likely
  present.}

{\noindent {\bf b)} The difference between approaching and receding
  jet must reflect the light travel time difference $\tau_{\rm light}
  \approx 2l_{\rm jet}\cot{i}/c \approx 10\,{\rm yrs}\,D_{7.8}\cot{i}$
  between both sides if the jet is propagating close to the speed of
  light. $\tau_{\rm light}$ is significantly longer than the
  variability time scale for state transitions.  Thus, it is plausible
  that the absence of the approaching \mbox{X-ray} jet reflects different
  activity levels of the source over roughly a 10 yr period.}

\subsection{Broader implications}
\label{sec:summary}
Chandra found extended \mbox{X-ray} jets for a number of either
identified or likely black hole \mbox{X-ray} binaries (XTE J1550-564,
\citealt{corbel:02}; SS433, \citealt{migliari:02}; H1743-322,
\citealt{corbel:05}; and Cygnus X-3, \citealt{heindl:03}). \cirx~is
thus the first bona fide neutron star with an extended \mbox{X-ray}
jet.

The minimum power of $W\gtrsim 5\times 10^{36}\,{\rm ergs\,s^{-1}}$ is
roughly 4\% of the Eddington luminosity for a neutron star. At peak
flux, \cirx~is only slightly super-Eddington, and because, unlike in
black holes, mass cannot disappear into an event horizon, we can turn
this into a lower limit on the jet production efficiency of $\eta
\gtrsim 0.5\%$, that is: upward of 0.5\% of the accreted rest mass
energy must have gone into powering the X-ray jet.  This makes
\cirx~as efficient as a black hole in making jets, despite its
shallower potential.  It also ties
\cirx~directly to the radio nebula, showing that the jets have more
than enough power to inflate the large scale lobes.

{\noindent \it Acknowledgments:} We would like to thank Mike Nowak
and John Davis for their insight and support.  NSS and WNB acknowledge
support through NASA grant G05-6040X.

\end{document}